\documentclass{JHEP3}
\usepackage{amssymb}

\def \be {\begin{equation}}
\def \ee {\end{equation}}

\def \bea {\begin{eqnarray}}
\def \eea {\end{eqnarray}}
\def \nn {\nonumber}

\def \a {\alpha}
\def \b {\beta}

\def \G {\Gamma}
\def \d {\delta}

\def \m {\mu}
\def \n {\nu}
\def \k {\kappa}

\def \s {\sigma}
\def \r {\rho}
\def \o {\omega}
\def \O {\Omega}
\def \th {\theta}
\def \Th {\Theta}

\def \t {\tau}
\def \dag {\dagger}
\def \p {\partial}

\def\bd{\begin{document}}
\def\ed{\end{document}}
\def\nn{\nonumber}
\def\bea{\begin{eqnarray}}
\def\eea{\end{eqnarray}}
\let\bm=\bibitem
\let\la=\label

\def\N{{\cal N}}
\def\sst{\scriptscriptstyle}
\def\thetabar{\bar\theta}
\def\Tr{{\rm Tr}}
\def\one{\mbox{1 \kern-.59em {\rm l}}}

\def\a{\alpha}      \def\da{{\dot\alpha}}
\def\b{\beta}       \def\db{{\dot\beta}}
\def\c{\gamma}  \def\C{\Gamma}  \def\cdt{\dot\gamma}
\def\d{\delta}  \def\D{\Delta}  \def\ddt{\dot\delta}
\def\e{\epsilon}        \def\vare{\varepsilon}
\def\f{\phi}    \def\F{\Phi}    \def\vvf{\f}
\def\h{\eta}
\def\k{\kappa}
\def\l{\lambda} \def\L{\Lambda}
\def\m{\mu} \def\n{\nu}
\def\o{\omega}
\def\P{\Pi}
\def\r{\rho}
\def\s{\sigma}  \def\S{\Sigma}
\def\t{\tau}
\def\th{\theta} \def\Th{\Theta} \def\vth{\vartheta}
\def\X{\Xeta}
\def\z{\zeta}
\def\w{\wedge}
\def\u{\underline}
\def\hs{\hspace}

\def\cA{{\cal A}} \def\cB{{\cal B}} \def\cC{{\cal C}}
\def\cD{{\cal D}} \def\cE{{\cal E}} \def\cF{{\cal F}}
\def\cG{{\cal G}} \def\cH{{\cal H}} \def\cI{{\cal I}}
\def\cJ{{\cal J}} \def\cK{{\cal K}} \def\cL{{\cal L}}
\def\cM{{\cal M}} \def\cN{{\cal N}} \def\cO{{\cal O}}
\def\cP{{\cal P}} \def\cQ{{\cal Q}} \def\cR{{\cal R}}
\def\cS{{\cal S}} \def\cT{{\cal T}} \def\cU{{\cal U}}
\def\cV{{\cal V}} \def\cW{{\cal W}} \def\cX{{\cal X}}
\def\cY{{\cal Y}} \def\cZ{{\cal Z}}
\def\bo {\bar{\o}}

\def\ua{\underline{\alpha}} \def\ubb{\underline{\beta}}
\def\ug{\underline{\gamma}}
\def\ub{\underline{\phantom{\alpha}}\!\!\!\beta}
\def\uc{\underline{\phantom{\alpha}}\!\!\!\gamma}
\def\um{\underline{\mu}} \def\un{\underline{\nu}}
\def\ud{\underline\delta}
\def\ue{\underline\epsilon}
\def\una{\underline a}\def\unA{\underline A}
\def\unb{\underline b}\def\unB{\underline B}
\def\unc{\underline c}\def\unC{\underline C}
\def\und{\underline d}\def\unD{\underline D}
\def\une{\underline e}\def\unE{\underline E}
\def\unf{\underline{\phantom{e}}\!\!\!\! f}\def\unF{\underline F}
\def\unm{\underline m}\def\unM{\underline M}
\def\unn{\underline n}\def\unN{\underline N}
\def\unp{\underline{\phantom{a}}\!\!\! p}\def\unP{\underline P}
\def\unq{\underline{\phantom{a}}\!\!\! q}
\def\unQ{\underline{\phantom{A}}\!\!\!\! Q}
\def\unH{\underline{H}}
\def\ul{\underline}
\def\As {{A \hspace{-6.4pt} \slash}\;}
\def\bs {{b \hspace{-6.4pt} \slash}\;}
\def\Ds {{D \hspace{-6.4pt} \slash}\;}
\def\ds {{\del \hspace{-6.4pt} \slash}\;}
\def\ss {{\s \hspace{-6.4pt} \slash}\;}
\def\ks {{ k \hspace{-6.4pt} \slash}\;}
\def\ps {{p \hspace{-6.4pt} \slash}\;}
\def\pas {{{p_1} \hspace{-6.4pt} \slash}\;}
\def\pbs {{{p_2} \hspace{-6.4pt} \slash}\;}

\def\Fh{\hat{F}}
\def\Vh{\hat{V}}
\def\Xh{\hat{X}}
\def\ah{\hat{a}}
\def\xh{\hat{x}}
\def\yh{\hat{y}}
\def\ph{\hat{p}}
\def\xih{\hat{\xi}}

\def\psit{\tilde{\psi}}
\def\Psit{\tilde{\Psi}}
\def\tht{\tilde{\th}}

\def\At{\tilde{A}}
\def\Qt{\tilde{Q}}
\def\Rt{\tilde{R}}
\def\Nt{\tilde{N}}

\def\at{\tilde{a}}
\def\st{\tilde{s}}
\def\ft{\tilde{f}}
\def\pt{\tilde{p}}
\def\qt{\tilde{q}}
\def\vt{\tilde{v}}
\def\nt{\tilde{n}}

\def\delb{\bar{\partial}}
\def\bz{\bar{z}}
\def\bD{\bar{D}}
\def\bB{\bar{B}}
\def\bo {\bar{\o}}

\def\bk{{\bf k}}
\def\bl{{\bf l}}
\def\bp{{\bf p}}
\def\bq{{\bf q}}
\def\br{{\bf r}}
\def\bx{{\bf x}}
\def\by{{\bf y}}
\def\bR{{\bf R}}
\def\bV{{\bf V}}
\def\bd{\begin{document}}

\def\ed{\end{document}}

\def\d{\delta}\def\D{\Delta}\def\ddt{\dot\delta}

\def\p{\partial} \def\del{\partial}
\def\xx{\times}
\def\uno{\mbox{1 \kern-.59em {\rm l}}}

\def\trp{^{\top}}
\def\inv{^{-1}}
\def\dag{{^{\dagger}}}
\def\pr{\prime}

\def\rar{\rightarrow}
\def\lar{\leftarrow}
\def\lrar{\leftrightarrow}

\def\cw{{\cal W}}
\def\cz{{\cal Z}}
\def\tcm{\tilde{\cal M}}
\def\sgn{{\rm sgn}}
\def\sd {d^{4|4}}
\def\lan{\langle}
\def\ran{\rangle}

\title{Real-time Correlators and Hidden Conformal Symmetry in the Kerr/CFT Correspondence}
\author{Bin Chen\\
Department of Physics,\\
and State Key Laboratory of Nuclear Physics and Technology,\\
and Center for High Energy Physics,\\
Peking University,\\
Beijing 100871, P.R. China\\
\email{bchen01@pku.edu.cn}}

\author{Jiang Long\\
Department of Physics,\\
Peking University,\\
Beijing 100871, P.R. China\\
\email{longjiang0301@gmail.com}}

\date{\today}
\abstract{In this paper, we study the real-time correlators in
Kerr/CFT, in the low frequency limit of generic non-extremal
Kerr(-Newman) black holes. From the low frequency scattering off
Kerr-Newman black holes, we show that for the uncharged scalar
scattering, there exists hidden conformal symmetry on the solution
space. Similar to Kerr case, this suggests that the Kerr-Newman
black hole is dual to a two-dimensional CFT with central charges
$c_L=c_R=12J$ and temperatures $T_L=\frac{(r_++r_-)-Q^2/M}{4\pi
a}, T_R=\frac{r_+-r_-}{4\pi a}$. Using the Minkowski prescription,
we compute the real-time correlators of a charged scalar and find
perfect match with CFT prediction. We further discuss the
low-frequency scattering of photons and gravitons by a Kerr black
hole and find that their retarded Green's functions are in good
agreement with CFT prediction. Our study shows that
 hidden conformal symmetry in the solution space is essential
to set up and check the Kerr/CFT correspondence.
 }

\newpage
\begin{document}

\section{Introduction}\label{sec-intro}

The original Kerr/CFT  correspondence conjectures that the quantum
gravity in the near-horizon extreme Kerr (NHEK) geometry with
certain boundary conditions is dual to a (1+1) dimensional chiral
conformal field theory (CFT). The correspondence was inspired by
the  properties of the asymptotic symmetry group of the near
horizon geometry \cite{Bardeen:1999px} of an extreme Kerr black
hole where it was found
%c2
by Guica, Hartman, Song and Strominger (GHSS) \cite{AndyWei} that
under a certain  set of  boundary conditions on the asymptotic
behaviour of the metric, the $U(1)_L$ symmetry of the  $SL(2,R)_R
\times U(1)_L$ isometry group \cite{Brown86} of the near-horizon
%c2 extreme Kerr NHEK
geometry get enhanced into a Virasoro algebra. Another copy of
Virasoro algebra was later discovered from the analysis of the
NHEK boundary condition in \cite{{Castro:2009jf}, matsuo}. Now it
is believed that the extreme Kerr black hole is dual to a CFT with
central charges $c_L=c_R=12J$. Support of this conjecture has been
found in  the perfect match of the macroscopic Berenstein-Hawking
entropy of the  black hole with the conformal field theory entropy
computed by the  Cardy formula. See  \cite{Lu:2008jk} for some
further studies  of the Kerr/CFT correspondence as well as
generalizations to other spacetime which contain a warped AdS
structure.

%c2 More recently,
Further support of the correspondence were found in the studies of
the superradiant scattering processes off the extreme Kerr black holes
%c2 are studied
\cite{Bredberg:2009pv}.
In this case, the near horizon geometry
of a near-extremal Kerr black hole
(near-NHEK) is reminiscent of a
non-extremal warped black hole. Correspondingly, the right-moving
sector of dual CFT is excited \cite{Castro:2009jf}. In the near-horizon limit, the
modes of interest are the ones near the super-radiant bound. It was shown in \cite{Bredberg:2009pv}
that the bulk scattering results
were in precise agreement with the CFT description whose form is
completely fixed by conformal invariance.
Similar discussion has been generalized to charged
Kerr-Newman \cite{Hartman:2009nz}, multi-charged
Kerr \cite{Cvetic:2009jn} and higher dimensional near-extremal Kerr
black holes. In all these cases, perfect agreements with the dual CFT
descriptions have been found.

In \cite{ChenChu}, it was shown that the real-time correlators of
various perturbations in near-extremal Kerr(-Newman) black hole
could be computed directly from the bulk, following the Minkowski
prescription proposed in AdS/CFT \cite{Son05} and successfully
used in the warped AdS/CFT correspondence\cite{Chen:2009cg}. It
allowed us  to perform a test directly on the CFT correlators and
the real-time correlators as obtained by holography. The results
are in perfect agreement with the CFT predictions. The similar
prescription has been applied to calculate the three-point
functions in the Kerr/CFT correspondence\cite{Becker:2010jj}.

It has been expected that the Kerr/CFT correspondence should be
true for general $J$ and $M$, which could be far away from extreme
limit. However, for a generic  non-extremal Kerr black holes, the
NHEK geometry disappears and it is not clear how to associate a
CFT from the near horizon geometry. Very recently, in a remarkable
paper \cite{Castro:2010fd} the authors argued that the existence
of conformal invariance in a near horizon geometry is not a
necessary condition, instead the existence of a local conformal
invariance in the solution space of the wave equation for the
propagating field is sufficient to ensure a dual CFT description.
The hidden $SL(2,R)\times SL(2,R)$ conformal symmetry acting on
the solution space of the wave function is only a local symmetry,
and  is broken by the periodic identification in the configuration
space. The spontaneous breaking of the conformal symmetry  is of
the form produced by finite temperatures $(T_L,T_R)$ in the dual
2D CFT. It was suggested that a generic non-extremal Kerr black
hole is dual to a 2D CFT with central charges $c_L=c_R=12J$ and
temperatures $T_L=M^2/2\pi J$ and $T_R=\sqrt{M^4-J^2}/2\pi J$.
Both the microscopic entropy counting and the low frequency scalar
scattering amplitude in the  near region support the picture. The
other related studies could be found in
\cite{Krishnan:2010pv,Chen:2010as,Wang:2010qv}.

In this paper, we would like to study the physical implication of
the hidden conformal symmetry on the real-time correlators.
Firstly we investigate the low-frequency scattering of the scalar
by a Kerr-Newman black hole. We find that for a uncharged
massless scalar scattering off the Kerr-Newman black hole, there is a
hidden $SL(2,R)\times SL(2,R)$ conformal symmetry acting on the
solution space of the radial wave function. In this case, the dual
CFT is of the central charges $c_L=c_R=12J$, but with the
temperatures \be T_L=\frac{r_++r_--Q^2/M}{4\pi a},
\hs{5ex}T_R=\frac{r_+-r_-}{4\pi a}. \ee We recover the macroscopic
Bekenstein-Hawking entropy from the Cardy formula. We discuss the
low-frequency massless scalar scattering off the black hole in the
near region $r<<1/\o$. We use the Minkowski prescription of the
AdS/CFT correspondence to calculate the real-time correlators for
the charged scalar and find perfect match with  the CFT prediction.
Then we turn to the low-frequency scattering of vector and
gravitational perturbations off a Kerr black hole. Using the
prescription proposed in \cite{ChenChu}, we compute the real-time
correlators for these perturbations and once again find perfect
agreement with the CFT prediction.

In the next section, we study the low frequency scalar scattering off the Kerr-Newman black hole.
In the near region, the wave function takes a form of hypergeometric function, suggesting a underlying
conformal invariance.
In section 3, we show the hidden $SL(2,R)\times SL(2,R)$ symmetry in the solution space of the massless uncharged scalar wave
function.  In
section 4,  we give the CFT description of the black hole entropy and finite-temperature absorption section of charged scalar, which are in perfect
match with the bulk results. In section 5,
  we discuss the low-frequency scattering of vector and gravitational perturbations off the Kerr
black hole. We compute the real-time correlators from the
Minkowski prescription, and find the agreement with the CFT
prediction. We end with some discussions in section 6.

{\bf Note added:} when we are finishing this project, we notice that there
appears a paper \cite{Wang:2010qv}, which overlaps with our results
in section 2 and 3.

\section{Scalar scattering of Kerr-Newman black hole}

For a Kerr-Newman black hole with mass $M$, angular momentum
$J=aM$ and electric charge $Q$, its metric takes the following form
\be \label{KerrNewman}
ds^2=-\frac{\D}{\r^2}(d t-a\sin^2\th d\phi)^2+
\frac{\r^2}{\D}dr^2+\rho^2 d\th^2+ \frac{1}{\r^2}\sin^2\th\left(ad
t-(r^2+a^2)d\phi\right)^2, \ee where \bea
\Delta&=&(r^2+a^2)-2Mr+Q^2, \nn\\
\r^2&=&r^2+a^2\cos^2\th.
\eea
%c2
The gauge field is \be A=-\frac{Qr}{\r^2}(d t-a\sin^2\th d\phi).
\ee There are two horizons located at \be r_\pm=M\pm
\sqrt{M^2-a^2-Q^2}. \ee And the Hawking temperature, entropy,
angular velocity of the horizon and the electric potential are
respectively
 \bea
 T_H&=&\frac{r_+-r_-}{4\pi(r_+^2+a^2)},\nn\\
 S&=&\pi(r_+^2+a^2),\nn\\
 \O_H&=&\frac{a}{r_+^2+a^2},\nn\\
 \Phi&=&\frac{Qr_+}{r_+^2+a^2}.
 \eea

Let us consider a complex scalar field with mass $\mu$ and charge $e$ scattering off
the Kerr-Newman black hole. The Klein-Gordon equation is
\be
(\nabla_\mu+ieA_\mu)(\nabla^\mu+ieA^\mu)\Phi-\mu^2\Phi=0.
\ee
With the ansatz
\be\label{ansatz}
 \Phi=e^{-i\o t+im\phi}\cR(r)\cS(\th),
 \ee
where $\o$ and $m$ are the quantum numbers, the wave equation could
be decomposed into the angular part and the radial part. The angular
part is of the form \be
 \frac{1}{\sin\th}\frac{d}{d\th}\left(\sin\th\frac{d}{d\th}\cS\right)+\left(
 \L_{lm}-a^2(\o^2-\mu^2)\sin^2\th-\frac{m^2}{\sin^2\th}\right)\cS=0.
 \ee
%c2 where
Here $\L_{lm}$ is the separation constant. It is
restricted by the regularity
boundary condition at $\th=0,\pi$ and can be computed
numerically. The radial part of the wave function is of the form
\be
\p_r(\D \p_r\cR)+V_R \cR=0
\ee
with
\bea
V_R&=&-\L_{lm}+2am\o+\frac{H^2}{\D}-\m^2(r^2+a^2), \\
H&=&\o(r^2+a^2)-eQr-am.
\eea

As we are interested in the low frequency limit, \be\label{low} \o
M <<1, \ee the $\o^2$ term in the angular equation could be
neglected. Note that the low frequency limit (\ref{low}) is very
different from the case studied in
\cite{Bredberg:2009pv,{Hartman:2009nz}}, where only the
frequencies near the superradiant bound were studied. To simplify
our discussion, we focus on the massless scalar. Then the angular
equation is just the Laplacian on the 2-sphere with the separation
constants taking values \be \L_{lm}=l(l+1). \ee

In the near region, $r\o <<1$, the radial equation could be simplified even more
\bea
\lefteqn{\p_r\D \p_r\cR(r)+\frac{(ma-\o(2Mr_+-Q^2)+eQr_+)^2}{(r-r_+)(r_+-r_-)}\cR(r)}\\
&&-\frac{(ma-\o(2Mr_--Q^2)+eQr_-)^2}{(r_+-r_-)(r-r_-)}\cR(r)=\left(l(l+1)-e^2Q^2\right)\cR(r)\label{radial}
\eea

Introduce
\be
z=\frac{r-r_+}{r-r_-},
\ee
we can rewrite the equation (\ref{radial}) as
\be
(1-z)z\p^2_z\cR(z)+(1-z)\p_z\cR(z)+\left(\frac{A_1}{z}+A_2+\frac{A_3}{1-z}\right)\cR(z)=0,\label{radial2}
\ee
where
\bea
A_1&=&\frac{(ma-\o(2Mr_+-Q^2)+eQr_+)^2}{(r_+-r_-)^2}, \nn\\
A_2&=&-  \frac{(ma-\o(2Mr_--Q^2)+eQr_-)^2}{(r_+-r_-)^2},\nn\\
A_3&=&-\left(l(l+1)-e^2Q^2\right)
\eea
The equation (\ref{radial2}) has the solution
\be
\cR(z)=z^\a (1-z)^\b F(a,b,c;z)
\ee
with
\be
\a=-i\sqrt{A_1},\hs{4ex}\b=\frac{1}{2}(1-\sqrt{1-4A_3}),
\ee
and
\bea
c&=&1+2\a, \\
a&=&\a+\b+i\sqrt{-A_2},\\
b&=&\a+\b-i\sqrt{-A_2}. \eea

If $\b$ is real, then in outer boundary of the matching region \be
M<<r<<\frac{1}{\o}, \ee the solution behaves asymptotically as
\be\label{chargedscalar} \cR(r) \simeq A r^{h-1}+B r^{-h} \ee where
$h$ is the conformal weight of the scalar field \be
h=1-\b=\frac{1}{2}(1+\sqrt{(2l+1)^2-4e^2Q^2}), \ee and \be
A=\frac{\G(c)\G(c-a-b)}{\G(c-a)\G(c-b)},
\hs{5ex}B=\frac{\G(c)\G(a+b-c)}{\G(a)\G(b)}. \ee

The absorption cross section could be read out
\bea
P_{abs}\sim |A|^{-2}&\sim & \sinh\left(2\frac{\o(2Mr_+-Q^2)-ma-eQr_+}{r_+-r_-}\right)|\G(h-i(2M\o-eQ)|^2 \nn\\
 &&\left|\G\left(h-i2\frac{\o(2M^2-Q^2)-ma-eQM}{r_+-r_-}\right)\right|^2. \label{crosssection}
 \eea

\section{Hidden conformal symmetry}

In this section, we will show that for the massless uncharged
particle, there exists a hidden $SL(2,R)_L \times SL(2,R)_R$
conformal symmetry acting on the solution space. Furthermore, from
the spontaneous breaking of this hidden symmetry by periodic
identification of $\phi$, we can read out  the left and right
temperatures of the dual conformal field theory.

From the coordinates
\bea
\o^+&=&\sqrt{\frac{r-r_+}{r-r_-}}e^{2\pi T_R\phi+2n_R t},\nn\\
\o^-&=& \sqrt{\frac{r-r_+}{r-r_-}}e^{2\pi T_L\phi+2n_L t},\nn\\
y&=&\sqrt{\frac{r_+-r_-}{r-r_-}}e^{\pi (T_L+T_R)\phi+(n_L+n_R)t},\nn
\eea
we can locally define the vector fields
\bea
H_1&=&i\p_+ \nn\\
H_0&=&i\left(\o^+\p_++\frac{1}{2}y\p_y\right) \nn\\
H_{-1}&=&i(\o^{+2}\p_++\o^+y\p_y-y^2\p_-)
\eea
and
\bea
\tilde H_1&=&i\p_- \nn\\
\tilde H_0&=&i\left(\o^-\p_-+\frac{1}{2}y\p_y\right) \nn\\
\tilde H_{-1}&=&i(\o^{-2}\p_-+\o^-y\p_y-y^2\p_+)
\eea
These vector fields obey the $SL(2,R)$ Lie algebra
\be
[H_0, H_{\pm 1}]=\mp iH_{\pm 1},\hs{5ex} [H_{-1},H_1]=-2iH_0,
\ee
and similarly for $(\tilde H_0, \tilde H_{\pm 1})$. The quadratic Casimir is
\bea
\cH^2=\tilde{\cH}^2&=&-H_0^2+\frac{1}{2}(H_1 H_{-1}+H_{-1}H_1) \nn\\
 &=&\frac{1}{4}(y^2\p^2_y-y\p_y)+y^2 \p_+\p_-.
 \eea

 In terms of $(t,r,\phi)$ coordinates, the Casimir becomes
 \bea
 \cH^2 &=& (r-r_+)(r-r_-)\frac{\p^2}{\p r^2}+(2r-r_+-r_-)\frac{\p}{\p r} \\
  &&+\frac{r_+-r_-}{r-r_-}\left(\frac{n_L-n_R}{4\pi A}\p_\phi -\frac{T_L-T_R}{4A}\p_t\right)^2
  -\frac{r_+-r_-}{r-r_+}\left(\frac{n_L+n_R}{4\pi A}\p_\phi
  -\frac{T_L+T_R}{4A}\p_t\right)^2\nn
  \eea
  where $A=n_LT_R-n_R T_L$. We find that
  with the following identification
  \bea
  n_R=0, & &T_R=\frac{r_+-r_-}{4\pi a} \nn\\
  n_L=-\frac{1}{4M}, & & T_L=\frac{(r_++r_-)-Q^2/M}{4\pi a}, \label{identification}
  \eea
 the radial equation (\ref{radial}) of neutral scalar, that is $e=0$, is the same as
 \be
 \tilde{\cH}^2\cR(r)=\cH^2 \cR(r)=l(l+1) \cR(r).
 \ee
 In other words, the scalar Laplacian is just the $SL(2,R)$ Casimir.

 As pointed out in \cite{Castro:2010fd}, the vector fields are not globally defined. In fact, the $SL(2,R)\times SL(2,R)$ symmetry
 is spontaneously broken down to $U(1)_L\times U(1)_R$ subgroup by the periodic identification
 \be
 \phi \sim \phi+2\pi.
 \ee

 Note that  the temperature identification in (\ref{identification}) reflects the nature of the underlying geometry. In the $Q\to 0$ limit,
 it reduces to the one in the Kerr case. It should be universal to
 all kinds of perturbations, including the charged scalar,  even
 though it was obtained from the radial equation of a massless neutral scalar.

 \section{Microscopic description}

  The Kerr/CFT correspondence suggests that the Kerr-Newman black hole is dual to a CFT with central charges
 \be
 c_L=c_R=12J   \label{central}
 \ee
 at finite temperature $(T_L, T_R)$ given in (\ref{identification}).  This should be true for every value of angular momentum and charge.

As a first check of this conjecture in the Kerr-Newman case, we show that the entropy of the black hole could be recovered from dual CFT.
The Cardy formula gives the microscopic entropy
\be
S=\frac{\pi^2}{3}(c_L T_L+c_R T_R).
 \ee
From the central charges (\ref{central}) and the temperatures  (\ref{identification}), we have
\be
S=\pi(r^2_++a^2)
\ee
which is in perfect agreement with the macroscopic Bekenstein-Hawking area law for the entropy of the Kerr-Newman black hole.

To determine the conjugate charges, we begin with the first law of
thermodynamics \be T_H \d S=\d M-\O \d J-\Phi \d Q. \ee We are
looking for the conjugate charges $\d E_L$ and $\d E_R$ such that
\be \d S=\frac{\d E_L}{T_L}+\frac{\d E_R}{T_R}. \ee The solution is
\bea
\d E_L&=&\frac{(2M^2-Q^2)M}{J}\d M-\frac{QM^2\d Q}{J}+\frac{Q^3\d Q}{2J}, \nn\\
\d E_R&=&\frac{(2M^2-Q^2)M}{J}\d M-\frac{QM^2\d Q}{J}-\d J, \eea If
we identify \bea
&&\d M=\o, \hs{3ex}\d J=m,\hs{3ex}\d Q=e, \nn\\
&&\o_L=\frac{(2M^2-Q^2)M}{J}\o, \hs{3ex}\o_R=\frac{(2M^2-Q^2)M}{J}\o-m, \label{identification1} \\
&&q_L=q_R=\d Q=e, \label{identification2}\\
&&\m_L=\frac{QM^2}{J}-\frac{Q^3}{2J},
\hs{3ex}\m_R=\frac{QM^2}{J}\label{identification3} \eea we have \be
\d E_L=\o_L-q_L\m_L, \hs{3ex}\d E_R=\o_R-q_R\m_R. \ee In other
words, if the charge $e$ is nonvanishing, it couples to both the
left and right chemical potential. If both $e$ and $Q$ are
vanishing, the above identifications reduce to the ones in the Kerr
case.

With the identification
(\ref{identification1}-\ref{identification3}), the absorption cross
section (\ref{crosssection}) could be rewritten as
\bea\label{scalarsection}
P_{abs}&\sim& T_L^{2h_L-1}T_R^{2h_R-1}\sinh\left(\frac{\o_L-q_L\m_L}{2T_L}+ \frac{\o_R-q_R\m_R}{2T_R}\right)\times \nn\\
& &\left|\G\left(h_L+i\frac{\o_L-q_L\m_L}{2\pi T_L}\right)\right|^2\cdot \left|\G\left(h_R+i\frac{\o_R-q_R\m_R}{2\pi T_R}\right)\right|^2
\eea

In a 2D conformal field theory(CFT), one can define the two-point
function \be
 G(t^+,t^-)=\langle {\cal O}^\dagger_\phi(t^+,t^-){\cal
 O}_\phi(0)\rangle,
\ee where $t^+,t^-$ are the left and right moving coordinates of
2D worldsheet and  ${\cal O}_\phi$ is the operator corresponding
to the field perturbing the black hole. For an operator of
dimensions $(h_L,h_R)$, charges $(q_L,q_R)$ at temperatures
$(T_L,T_R)$ and chemical potentials $(\m_L, \m_R)$, the two-point
function  is dictated by conformal invariance and takes the form
\cite{Cardy:1984bb}:
 \be \label{G-Mink}
 G(t^+,t^-)\sim (-1)^{h_L+h_R}\left(\frac{\pi T_L}{\sinh(\pi T_L
 t^+)}\right)^{2h_L}\left(\frac{\pi T_R}{\sinh(\pi T_R
 t^-)}\right)^{2h_R}e^{iq_L\m_L t^+ +iq_R\m_R t^-}.
 \ee
The CFT absorption cross
 section could be defined with the two-point functions, following
 Fermi's golden rule:
 \be
 \s_{abs}\sim \int dt^+dt^-
 e^{-i\o_Rt^--i\o_Lt^+}[G(t^+-i\epsilon,
  t^--i\epsilon)-G(t^++i\epsilon, t^-+i\epsilon)]
 \ee
Then after being changed into momentum space, the absorption cross
section is
 \bea\label{2Dsection}
 \s &\sim&
 T_L^{2h_L-1}T_R^{2h_R-1}\sinh\left(\frac{\o_L-q_L\m_L}{2T_L}+\frac{\o_R-q_R\m_R}{2T_R}\right)\nn\\
 && \left|\G\left(h_L+i\frac{\o_L-q_L\m_L}{2\pi T_L}\right)\right|^2\left|\G\left(h_R+i\frac{\o_R-q_R\O_R}{2\pi
 T_R}\right)\right|^2.
 \eea
 Obviously the absorption cross section of a charged scalar in Kerr-Newman black hole
 (\ref{scalarsection}) is in perfect match with the prediction of dual CFT.
%c1
%c1 I do not understand the following statements:
%c1  Actually, it would be appropriate to take the global geometry
%c1 (\ref{global}) as the vacuum, which allow us to bypass the claim
%%c1 that there is no dynamics  in extremal Kerr throat \cite{amsel}.
%c1 In this sense, both NHEK and near-NHEK
%c1 are black holes asymptotic to the global geometry.

Actually, in the Kerr/CFT correspondence, we can do slightly better.
Besides the absorption cross section, we may compare the retarded
correlators in  the bulk with the Euclidean correlators in dual
CFT.

 The Euclidean correlator $G_E$ is obtained
by a Wick rotation $t^+ \to i \t_L$, $t^- \to i \t_R$. At finite
temperature the Euclidean time is taken to have period $2 \pi/T_L,
2 \pi/T_R$ and  the momentum space Euclidean correlator is given
by \be G_E(\o_{L,E}, \o_{R,E}) = \int_{0}^{2\pi/T_L} d\t_L
\int_{0}^{2\pi/T_R} d \t_R\; e^{-i \o_{L,E}\t_L - i\o_{R,E}\t_R}
G_E(\t_L,\t_R), \ee where the Euclidean frequencies are related to
the Minkowskian ones by \be \o_{L,E} = i \o_L, \qquad \o_{R,E} = i
\o_R. \ee The integral is divergent but can be defined by analytic
continuation, one obtains \cite{Maldacena:1997ih}
 \bea \label{GE}
G_E(\o_{L,E}, \o_{R,E}) &\sim& T_L^{2 h_L-1}  T_R^{2 h_R-1} e^{i
\frac{\bo_{L,E}}{2T_L}} e^{i \frac{\bo_{R,E}}{2T_R}}\nn\\
&&\cdot\G(h_L + \frac{\bo_{L,E}}{2 \pi T_L})\G(h_L -
\frac{\bo_{L,E}}{2 \pi T_L}) \G(h_R + \frac{\bo_{R,E}}{2 \pi
T_R})\G(h_R - \frac{\bo_{R,E}}{2 \pi T_R}), \eea where \be
\bo_{L,E}= \o_{L,E} - i q_L \m_L, \quad \bo_{R,E}= \o_{R,E} - i
q_R \m_R. \ee

The retarded correlator  $G_R (\o_L, \o_R)$ is analytic on the
upper half complex $\o_{L,R}$-plane and its value along the
positive imaginary $\o_{L,R}$-axis gives the Euclidean correlator:
\be \label{GER} G_E(\o_{L,E}, \o_{R,E}) = G_R(i\o_{L,E},
i\o_{R,E}), \quad \o_{L,E} , \o_{R,E} >0. \ee This relation holds
both for zero and finite temperature. However
%c2
at finite temperature, $\o_{L,E}$ and $\o_{R,E}$ take discrete
values of the Matsubara frequencies \be \o_{L,E} =  2 \pi m_L T_L,
\quad \o_{R,E} =  2 \pi m_R T_R, \ee where $m_L, m_R$ are integers
for bosonic modes and are half integers for fermionic modes.

In \cite{ChenChu}, it was showed that the real-time correlators
could be computed holographically in the bulk. This is feasible
because that although the NHEK geometry is more complicated, it is
in fact a warped AdS${}_3$  spacetime with a warping factor being
a function of the angular variable. Therefore one can consider the
Kerr/CFT correspondence as a generalization of the warped AdS/CFT
correspondence.  However, as  the near horizon geometry of generic
non-extremal Kerr(-Newman) black holes are far from the warped AdS
or AdS spacetime, it is not certain if we can still apply the
prescription to compute the real-time correlators. Nevertheless,
the existence of the hidden  conformal invariance in the solution
space gives us confidence that the Minkowski prescription could be
applied even for generic non-extremal Kerr(-Newman) black holes.
We now show this is the case.

For a scalar field in a black hole background, the prescription for
two-point real-time correlators was first proposed in \cite{Son05}.
It could be simplified as follows. For the scalar wave function
satisfying the ingoing boundary condition at the black hole horizon,
its asymptotic behavior is \be \phi \sim A r^{h-1}+B r^{-h}. \ee
Then taken $A$ as the source term and $B$ as the response term, the
two-point retarded correlator is just \bea G_R &\sim & \frac{B}{A}
\eea

For the charged scalar scattering off the Kerr-Newman black hole,
its asymptotic behavior looks like (\ref{chargedscalar}. Therefore
its retarded Green's function is just
 \bea
G_R&\sim&\frac{\G(1-2h)}{\G(2h-1)}\frac{\G\left(h+i\frac{\o_L-q_L\m_L}{2\pi
T_L}\right)\G\left(h+i\frac{\o_R-q_R\m_R}{2\pi T_R}\right)}
 {\G\left(1-h+i\frac{\o_L-q_L\m_L}{2\pi T_L}\right)\G\left(1-h+i\frac{\o_R-q_R\m_R}{2\pi
 T_R}\right)}
  \eea
 with the identification (\ref{identification1}-\ref{identification3}). This
is in good match with the CFT prediction (\ref{GER}).

%c1 \section{Green's functions in 2D CFT}
\section{Photons and gravitons in Kerr black hole}

In this section, we study the low-frequency scattering of
electromagnetic and gravitational perturbation off a Kerr black hole.
The metric of the Kerr black hole could be obtained by simply
setting $Q=0$ in (\ref{KerrNewman}). To study the perturbations with
nonvanishing spin,  one has to apply the Newman-Penrose formalism
\cite{Newman:1961qr}. The problem has been well studied in
\cite{Teukolsky:1973ha,Press:1973zz,Teukolsky:1974yv,Starobinsky1,{Starobinsky2}}.
It turned out that the equations of motion of the
 perturbations
%c2 could
can be decomposed into two separated equations of
 motion. The wave function could be decomposed into the form
  \be
  \Psi^s=e^{-i\o t+im\phi}\cR^s(r)\cS^s(\th).
  \ee
$\Psi^s$ are related to the electromagnetic field strength and Weyl tensor for spin-1 and spin-2 perturbations.
The angular and radial functions satisfy the Teukolsky equations:
\be \label{angular}
 \frac{1}{\sin\th}\frac{d}{d\th}\left(\sin\th\frac{d}{d\th}\cS^s(\th)\right)
+\left(
 \L^s_{lm}-a^2\o^2\sin^2\th-2a\o s\cos\th
-\frac{m^2+s^2+2ms\cos\th}{\sin^2\th}\right)\cS^s(\th)=0,
 \ee
 \be
 \D^{-s}\frac{d}{dr}\left(\D^{s+1}\frac{d}{dr}\cR^s\right)+\left(\frac{H^2-2is(r-M)H}{\D}+4is\o r+2am\o+s(s+1)-\L^s_{lm}\right)\cR^s=0,
 \ee
 where in the Kerr case
 \bea
 \D&=& (r^2+a^2)-2Mr \nn\\
 H&=&\o(r^2+a^2)-am.
 \eea
 Here $\L^s_{lm}$ is the separation constant, depending on
 $l,m,s$ and satisfying
 \be\label{Lam}
 \L^s_{lm}(s)=\L^s_{lm}(-s).
 \ee

  In the low-frequency limit $\o <<1/M$, the $\o$ dependent terms in the angular equation could be neglected.
As a result the separation constant is only a function of $l,m,s$.
In the near region $r \o <<1$, the radial equation reduces to \be
\D^{-s}\frac{d}{dr}\left(\D^{s+1}\frac{d}{dr}\cR^s\right)+V_s
\cR^s=0 \ee with \be
V_s=\frac{\s_+^2-2is(r_+-M)\s_+}{(r-r_+)(r_+-r_-)}-
\frac{\s_-^2-2is(r_--M)\s_-}{(r-r_-)(r_+-r_-)}+s(s+1)-\L^s_{lm}, \ee
where
 \be
 \s_+=2M\o r_+-am, \hs{5ex}\s_-=2M\o r_--am
 \ee
The solution satisfying the ingoing boundary condition at the
horizon could be once again written in terms of hypergeometric
function
 \be
 \cR=z^\a (1-z)^\b F(a,b,c;z),
 \ee
 where $z=\frac{r-r_+}{r-r_-}$ and
 \bea
 \a&=&-i\frac{\s_+}{r_+-r_-} \nn\\
 \b&=& s+\frac{1}{2}(1+\sqrt{1+4\L^s_{lm}}) \nn\\
 a&=&\frac{1}{2}(1-\sqrt{1+4\L^s_{lm}})-2i\frac{2M^2\o-ma}{r_+-r_-}
 \nn\\
 b&=&s+\frac{1}{2}(1-\sqrt{1+4\L^s_{lm}})-i(2M\o)\nn\\
 c&=&1+s-i\frac{2\s_+}{r_+-r_-}.
 \eea

 The asymptotic behavior of the radial wave function is
 \be
 \cR^s(r) \sim A^s r^{h-1-s}+B^s r^{-h-s},
 \ee
 where
 \bea
 h^s&=&\frac{1}{2}(1+\sqrt{1+4\L^s_{lm}})\nn\\
 A^s&=&\frac{\G(\sqrt{1+4\L^s_{lm}})}{\G(s+h^s-i(2M\o))\G\left(h^s-2i\frac{2M^2\o-ma}{r_+-r_-}\right)}\nn\\
 B^s&=&\frac{\G(-\sqrt{1+4\L^s_{lm}})}{\G(s+1-h^s-i(2M\o))\G\left(1-h^s-2i\frac{2M^2\o-ma}{r_+-r_-}\right)}\nn
\eea One may calculate the absorption cross sections following the
way in \cite{Hartman:2009nz} and compare the results with the CFT
prediction. It turns out to be in perfect match. We will not present
the details here. Instead,  we  give an alternative derivation from
the retarded Green's functions.

For the vector and gravitational perturbations, the prescription has
been proposed in \cite{ChenChu}. If the radial wave function of the
perturbation with the spin $s$ satisfying the ingoing boundary
condition at the black hole horizon has asymptotic behavior as \be
\cR^s (r) \sim A^s r^{h-1-s}+B^s r^{-h-s}, \ee then the retarded
Green's function could be \be\label{pres}
 G_R^s \sim \frac{B^{-s}}{A^{s}}.
 \ee
 In our case, this leads to
 \be
 G^s_R \sim \frac{\G(1-2h^s)}{\G(2h^s-1)}\frac{\G\left(-s+h^s-i\frac{\o_L}{2\pi T_L}\right)\G\left(h^s-i\frac{\o_R}{2\pi T_R}\right)}
 {\G\left(s+1-h^s-i\frac{\o_L}{2\pi
 T_L}\right)\G\left(1-h^s-i\frac{\o_R}{2\pi T_R}\right)},
  \ee
  where we have used the identification (\ref{identification1}) with
  $Q=0$. Note that in Kerr case, the chemical potentials $\mu_{L,R}$ are absent.
With the conformal weights of the fields being identified as
 \be
 h^s_R=h^s, \hs{5ex}h^s_L=h^s_R-s,
 \ee
 the above retarded Green's function agrees precisely, up to a
 normalization factor, with the CFT result (\ref{GE}) at the
 Matsubara frequencies. The cross section can be read directly from
 the above Green's function
 \bea
 \s^s &\sim& {\rm Im}(G^s_R) \nn\\
   &\sim & \frac{1}{(\G(h^s_R-1))^2}\sinh\left(\frac{\o_L}{2T_L}+ \frac{\o_R}{2T_R}\right)\times \nn\\
& &\left|\G\left(h^s_L+i\frac{\o_L}{2\pi T_L}\right)\right|^2\cdot
\left|\G\left(h^s_R+i\frac{\o_R}{2\pi T_R}\right)\right|^2. \eea
They agree with the CFT result. The similar result has been obtained in the case
of four-dimensional black holes in string theory\cite{Cvetic:1997ap}.

\section{Discussions}

In this paper, we showed that there existed a hidden conformal
invariance in the low-frequency scattering off the Kerr-Newman black
holes. Even though the conformal symmetry is broken by periodic
identification in the configuration space, it acts on the solution
space and associate a dual CFT description to the black hole. More
precisely, the conformal coordinate transformation suggests that
the generic 4D Kerr-Newman black hole is dual to a 2D CFT with
central charges $c_L=c_R=12J$ and temperatures
$T_L=\frac{(r_++r_-)-Q^2/M}{4\pi a}, T_R=\frac{r_+-r_-}{4\pi a}$.
The Bekenstein-Hawking entropy could be recovered from the Cardy
formula counting the miscrostate degeneracy in the dual CFT. For the
charged scalar scattering, we identified the dual operators with
conformal dimension, left and right charges and chemical
potentials, which allow us to find perfect match with the CFT
prediction. Furthermore, we discussed the retarded two-point
correlators of vector and gravitational perturbations in the Kerr
black hole spacetime. Using the prescription suggested in
\cite{ChenChu}, we computed the retarded Green's functions and
found perfect agreement with the CFT Euclidean correlators, which
are restricted by conformal invariance. These agreements support
the belief that the conformal symmetry in the solution space could
be essential to set up a CFT description.

Note that the real-time correlator is closely related to the absorption cross section.
Actually the imaginary part of the retarded Green's function gives the greybody factor. On the other hand, the greybody factor could determine the retarded Green's function uniquely via spectral theorem\cite{Hartman:2009nz}. Therefore, the real-time correlators and the greybody factors should not be taken as independent check of the Kerr/CFT correspondence. Nevertheless, it is nice to see that one can apply the Minkowski prescription of AdS/CFT to obtain the  real-time correlators and find good agreement with the CFT predictions.

The fact that the conformal symmetry in the solution space is
sufficient to associate a CFT description have profound
implications. It may change our viewpoint that some kind of
AdS/CFT correspondence should rely on the geometry of the bulk
spacetime. It could happen in some parameter region, the solution
space has an enhanced conformal symmetry, suggesting a dual CFT
description. It would be interesting to understand this picture in
other situations, besides the Kerr/CFT correspondence.

The analysis in this paper and other related studies have focused  on
the low-frequency scattering limit. As it is well known, the dual
CFT description has been set up in the near-extremal limit, in which
case the focus of the frequency is near the super-radiant bound.
These two descriptions are consistent: the same central charges
suggesting the same CFT, consistent temperatures, even though the
conformal weights of the operators are slightly different. It is
interesting to see if we have the same picture for intermediate
frequencies.

Our computation focused on the retarded Green's functions. It could
be generalized to the three-point functions straightforwardly,
following the recipe in \cite{Becker:2010jj}.

The radial wave functions in all the cases we discussed are of
hypergeometric functions. This fact suggests that there could be hidden
conformal symmetry acting on the solution spaces of charged scalar,
vector and gravitational perturbations. It would be interesting to
work them out.

\section*{Acknowledgments}

 The work of BC was partially supported by NSFC Grant
No.10775002,10975005, and NKBRPC (No. 2006CB805905).

\end{document}